\documentclass[
,final
]{aipproc}
\layoutstyle{6x9}

\begin{document}

\title{Prospects for Low-Energy Antiproton Physics at Fermilab
}

\author{Daniel M. Kaplan}{address={Illinois Institute of Technology, Chicago, Illinois 60616, USA}}

\begin{abstract}
Fermilab has long had the world's most intense antiproton source. Despite this, 
opportunities for low-energy antiproton physics  at Fermilab have in the past been limited and\,---\,with the antiproton source now exclusively dedicated to serving the needs of the Tevatron Collider\,---\,are currently nonexistent. 
While the future of antiproton physics at Fermilab is
 uncertain, the anticipated shutdown of the Tevatron in about 2009 presents the opportunity for a world-leading low-energy antiproton program. We summarize
the current status of the Fermilab antiproton facility and review some current topics in hyperon physics
as examples of what might be achievable.
\end{abstract}

\maketitle

Fermilab has the world's highest-intensity antiproton source, and (thanks to the ongoing efforts of the Antiproton Source Dept.) the beam intensity it can provide continues to increase. 
With the planned shutdown of the Tevatron starting in 2009, the antiproton facility might then once again become available for low-energy studies. There is an extensive list of interesting particle-physics topics that can be addressed with such a facility~\cite{PANDA-TPR}, including ``unfinished business'' from the former LEAR and Fermilab low-energy antiproton programs. These include 
\begin{itemize}
\item 
precision $\overline{p}p\to\,$charmonium studies, begun by Fermilab E760 and E835; 
\item 
open-charm studies, including searches for $D/{\overline D}$ mixing and {\em CP} violation; 
\item
studies of  $\overline{p}p\to\,$hyperons, including searches for hyperon {\em CP} violation and rare decays; 
\item
the search for glueballs and gluonic hybrid states predicted by QCD; 
and
\item
trapped-$\overline{p}$ and antihydrogen studies.
\end{itemize}
Due to their requirements for beam energy (see Table~\ref{tab:thresh}) or intensity, all but the last of these cannot be done at the CERN Antiproton Decelerator. Many of them 
have been discussed 
in the context of the GSI FLAIR (Facility for Low-Energy Antiproton and Ion Research) project~\cite{FLAIR} and its general-purpose $\overline{\rm {P}}$ANDA detector~\cite{PANDA-TPR}. However, that facility is expected to have insufficient luminosity for the hyperon-physics topics just mentioned. 

\begin{table}[htb]
\caption{Thresholds for some processes of interest.}
\label{tab:thresh}
\begin{tabular}{lccc}
\hline\hline
Process &Threshold: & $\sqrt{s}$ [GeV]& momentum [GeV/$c$]\\
\hline\hline
$\overline{p}p\to\overline{\Lambda}\Lambda$  && 2.231 & 1.437\\
$\overline{p}p\to\overline{\Sigma}{}^-\Sigma^+$  && 2.379 & 1.854\\
$\overline{p}p\to\overline{\Xi}{}^+\Xi^-$ && 2.642 & 2.620 \\
$\overline{p}p\to\overline{\Omega}{}^+\Omega^-$ && 3.345 & 4.938\\
\hline
$\overline{p}p\to\eta_c$ && 2.980 & 3.678\\
$\overline{p}p\to\psi(3770)$ && 3.770 & 6.572  \\
\hline\hline
\end{tabular}
\end{table}

\section{Capabilities of the Fermilab Antiproton Source}

Fermilab's Antiproton Source (which includes the Debuncher ring and the Antiproton Accumulator ring, in which stochastic cooling is performed) now  cools and accumulates antiprotons at a maximum 
stacking rate
of $\approx$20\,mA/hr.  Given the 474\,m circumference of the  Accumulator, this corresponds to a production rate of 
$2\times10^{11}\,$antiprotons/hr and can thus
support a maximum luminosity of about $5\times10^{32}\,{\rm cm}^{-2}{\rm
s}^{-1}$ (i.e., beyond this luminosity, collisions would consume antiprotons more rapidly than they are produced). Fermilab's Main Injector project included construction (in the Main Injector tunnel) of the (permanent-magnet) Recycler ring,\footnote{So called because of its possible use to recycle antiprotons from a just-completed ${\overline p}p$ store for use in the next store; however, the Recycler ring enhances operation of the Antiproton Source in a number of ways: for example, it improves the performance of stochastic cooling as described in the text.} now being put into operation.
With the planned use of electron cooling in the Recycler, the stacking rate is likely to double by 2009. As discussed below,  $\sim10^{33}\,$cm$^{-2}$s$^{-1}$ luminosity is required for competitive reach in hyperon physics. 

The integrated luminosity of an experiment in the Accumulator could be limited by the 
beam-transfer capabilities of the complex: antiprotons can currently be transferred from the Accumulator to the Recycler, and from the Recycler to the Main Injector, but not vice versa. Since the stacking rate decreases approximately linearly with increasing store size, to maintain rapid stacking, beam is 
typically transferred to the Recycler when the Accumulator current reaches $\approx$100\,mA. 
The addition of a reverse-transfer beamline would allow stacked beam from the Recycler to be injected back into the Accumulator for collisions 
and could thus enhance the deliverable luminosity 
by perhaps a factor of 2.
Also desirable  is the ability to stack simultaneously with 
experimental running, which could double the efficiency of operation. This  could be provided by the addition of a small storage ring, which might be a new, custom-designed ring or an existing one, such as that from the Indiana University Cyclotron Facility or the CELSIUS ring from Uppsala University. Such a ring could also enhance the ability to decelerate antiprotons for trapping.

To understand some of the issues for a future low-energy antiproton facility
(in particular, the need for $10^{33}$ luminosity), it is useful to consider some 
physics examples.

\section{Hyperon {\em CP} violation}

In addition to that in $K$- and $B$-meson decay~\cite{Rosner}, the Standard
Model predicts {\em CP} violation in decays of
hyperons~\cite{Hyperon-CP,ACP,Valencia}.  The most accessible signals are differences
between the angular distributions of polarized-hyperon decay products and those of the corresponding antihyperons~\cite{ACP}. Precision measurement thus requires accurate knowledge of the polarizations of the
initial hyperons and antihyperons.

Angular-momentum conservation requires the final state in the decay of a spin-1/2 hyperon to a
spin-1/2 baryon plus a pion to be either S- or P-wave. 
As is well known, 
interference between the S- and P-wave
amplitudes causes parity violation, parametrized by Lee and
Yang~\cite{Lee-Yang} via two independent parameters $\alpha$ and
$\beta$ (proportional respectively to the real and imaginary parts
of the interference term). {\em CP} violation can arise as a difference in
$|\alpha|$ or $|\beta|$ between a hyperon decay and its {\em CP}-conjugate
antihyperon decay or as a particle-antiparticle difference in the partial
widths for such decays~\cite{ACP,Donoghue-etal}.

Table~\ref{tab:HCP} summarizes the experimental situation.  
The first three experiments cited studied
$\Lambda$ decay only~\cite{R608,DM2,PS185}, setting limits on the 
{\em CP} asymmetry parameter~\cite{ACP,Donoghue-etal}
\begin{equation}
A_{\Lambda}\equiv \frac{\alpha_{\Lambda}+
\alpha_{\overline{\Lambda}}}{\alpha_{\Lambda}-
\alpha_{\overline{\Lambda}}}\,,
\end{equation}
where $\alpha_\Lambda$ ($\alpha_{\overline{\Lambda}}$) characterizes the
$\Lambda$ ($\overline{\Lambda}$) decay  to  (anti)proton plus charged pion
 and, if 
 {\em CP} is conserved, $\alpha_\Lambda =
-\alpha_{\overline{\Lambda}}$. 

Fermilab experiments 756~\cite{E756} and 871 (``HyperCP'')~\cite{Holmstrom} and CLEO~\cite{CLEO} used $\Xi^-$ (${\overline \Xi}{}^+$) decay to produce polarized $\Lambda$'s, in whose subsequent decay the
slope of the (anti)proton angular distribution in the ``helicity'' frame 
measures the product $\alpha_\Xi\alpha_\Lambda$; for {\em
CP} conservation this should be identical for $\Xi$ and
$\overline{\Xi}$ events. The {\em CP} asymmetry parameter measured 
is thus 
\begin{equation}
A_{\Xi\Lambda}\equiv \frac{\alpha_{\Xi}\alpha_{\Lambda}-
\alpha_{\overline{\Xi}}\alpha_{\overline{\Lambda}}}{\alpha_{\Xi}\alpha_{\Lambda}+
\alpha_{\overline{\Xi}}\alpha_{\overline{\Lambda}}}
\approx A_\Xi + A_\Lambda\,.
\end{equation}
The power of this technique derives from the large $\alpha$ value ($\alpha=0.64$) in the
$\Xi\to\Lambda\pi$ decay. 
Also, in the fixed-target case, for a given 
${}^{{}^(\!}\overline{\Xi}{}^{{}^)}\!\!$-momentum bin the acceptances and
efficiencies for $\Xi$
and $\overline{\Xi}$ decays 
are very similar: Between
$\Xi$ and $\overline{\Xi}$ runs one reverses magnet polarities, making the spatial distributions of decay products across the detectors almost identical for $\Xi$ and  $\overline{\Xi}$. 
(There are still residual systematic uncertainties arising from the differing momentum dependences of the $\Xi$ and $\overline{\Xi}$ production cross sections and  of the cross sections for the $p$ and $\overline{p}$ and $\pi^+$ and $\pi^-$ to interact in the material of the spectrometer.)

HyperCP took data during 1996--99, recording the
world's largest samples of hyperon decays ($2.0 \times 10^9$ $\Xi$ and $4.6 \times 10^8$ $\overline\Xi$ events), and has set  the world's best limit on hyperon {\em CP} violation~\cite{Holmstrom}, based on about 5\% of the recorded data sample. The complete analysis should determine $A_{\Xi\Lambda}$ with a statistical uncertainty 
$\delta A = 
\sqrt{
{3}/{N_{\Xi^-}}+
{3}/{N_{\overline{\Xi}{}^+}}}/{2\alpha_{\Xi}\alpha_{\Lambda}}
 \stackrel{<}{_\sim} 2.0\times10^{-4}\,.
$ 
The Standard Model predicts $A_{\Xi\Lambda}\sim10^{-5}$~\cite{ACP}.   
Thus any significant effect seen in HyperCP will be evidence for {\em CP}
violation in the baryon sector substantially larger than predicted by the
Standard Model. Various Standard Model extensions predict effects as large as ${\cal O}(10^{-3})$~\cite{non-SM}.

\begin{table}
\caption {Summary of experimental limits on {\em CP} violation in hyperon decay.}
\label{tab:HCP}
\begin{tabular}{lccccc}
\hline\hline
Experiment & Facility & Year & Ref. & Mode & $A_\Lambda$ [$^*$] or $A_{\Xi\Lambda}$ [$^\dagger$] \\
\hline\hline
R608 & ISR & 1985 & \cite{R608} & $pp\to\Lambda X, pp\to\overline{\Lambda} X$ &
$-0.02\pm0.14^*$ \\
DM2 & Orsay & 1988 &  \cite{DM2} & $e^+e^- \to J/\psi \to \Lambda\overline{\Lambda}$ & 
$0.01\pm0.10^*$
\\
PS185 & LEAR & 1997 & \cite{PS185} & $\overline{p}p\to\overline{\Lambda}\Lambda$ & 
$0.006\pm0.014^*$ \\
& & & &
$e^+e^-\to\Xi^- X, \Xi^-\to\Lambda\pi^-,$ &\\
\raisebox{1.5ex}[0pt]{CLEO} & \raisebox{1.5ex}[0pt]{CESR} &\raisebox{1.5ex}[0pt]{2000} &
\raisebox{1.5ex}[0pt]{\cite{CLEO}} & $e^+e^-\to\overline{\Xi}{}^+ X, \overline{\Xi}{}^+\to\overline{\Lambda}\pi^+$ & 
\raisebox{1.5ex}[0pt]{$-0.057\pm0.064\pm
0.039^\dagger$}\\ 
&  & & &
$pN\to\Xi^- X, \Xi^-\to\Lambda\pi^-$, &  \\
\raisebox{1.5ex}[0pt]{E756} &\raisebox{1.5ex}[0pt]{Fermilab} & \raisebox{1.5ex}[0pt]{2000} &
\raisebox{1.5ex}[0pt]{\cite{E756}} & $pN\to\overline{\Xi}{}^+ X, \overline{\Xi}{}^+\to\overline{\Lambda}\pi^+$ & 
\raisebox{1.5ex}[0pt]{$0.012
\pm0.014^\dagger$} \\
&  & & &
$pN\to\Xi^- X, \Xi^-\to\Lambda\pi^-$, &  \\
\raisebox{1.5ex}[0pt]{HyperCP} &
\raisebox{1.5ex}[0pt]{Fermilab} & \raisebox{1.5ex}[0pt]{2004} &  \raisebox{1.5ex}[0pt]{\cite{Holmstrom}} &
$pN\to\overline{\Xi}{}^+ X, \overline{\Xi}{}^+\to\overline{\Lambda}\pi^+$ & 
\raisebox{1.5ex}[0pt]{$(0.0\pm5.1
\pm4.4)\times10^{-4}{}^{\,\dagger,\ddag}$} \\
\hline\hline
\end{tabular}

\footnotetext{$\ddag$ Based on $\approx$5\% of the HyperCP data sample; analysis of the full sample is still in progress.}
\end{table}

\section{Study of FCNC  $\Sigma^+$ decay}

\begin{figure}[htb]
\smallskip\leftline{\hspace{1.25in}
\begin{minipage}{0.5\linewidth}
\centerline{\includegraphics[width=\linewidth]{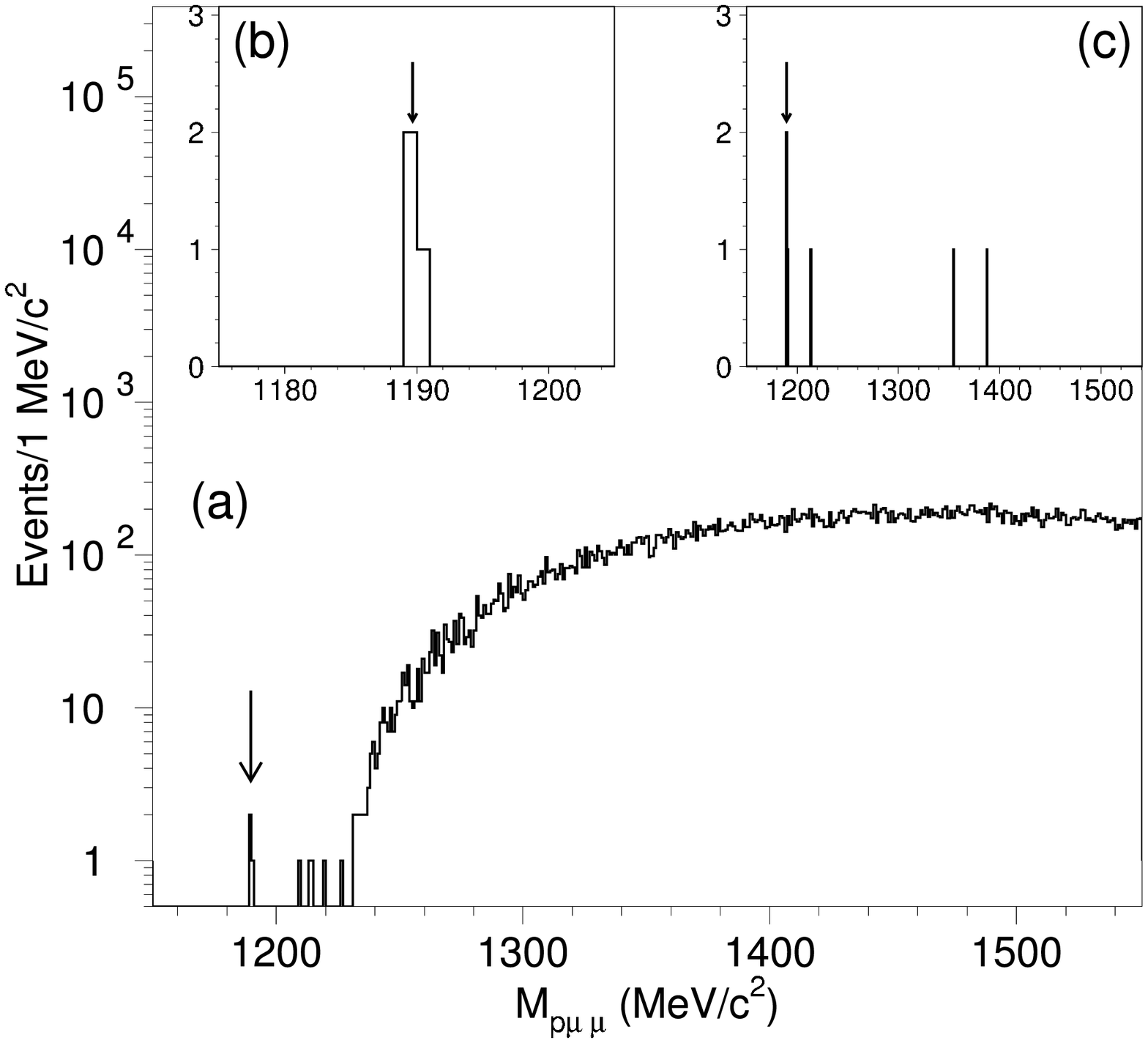}}\hspace{-.75in}
\label{fig:sigpmumu}
\end{minipage}}
\begin{minipage}{0.25\linewidth}
\hspace{-1.5in}
\centerline{\includegraphics[bb=40 100 500 650,width=3.1in]{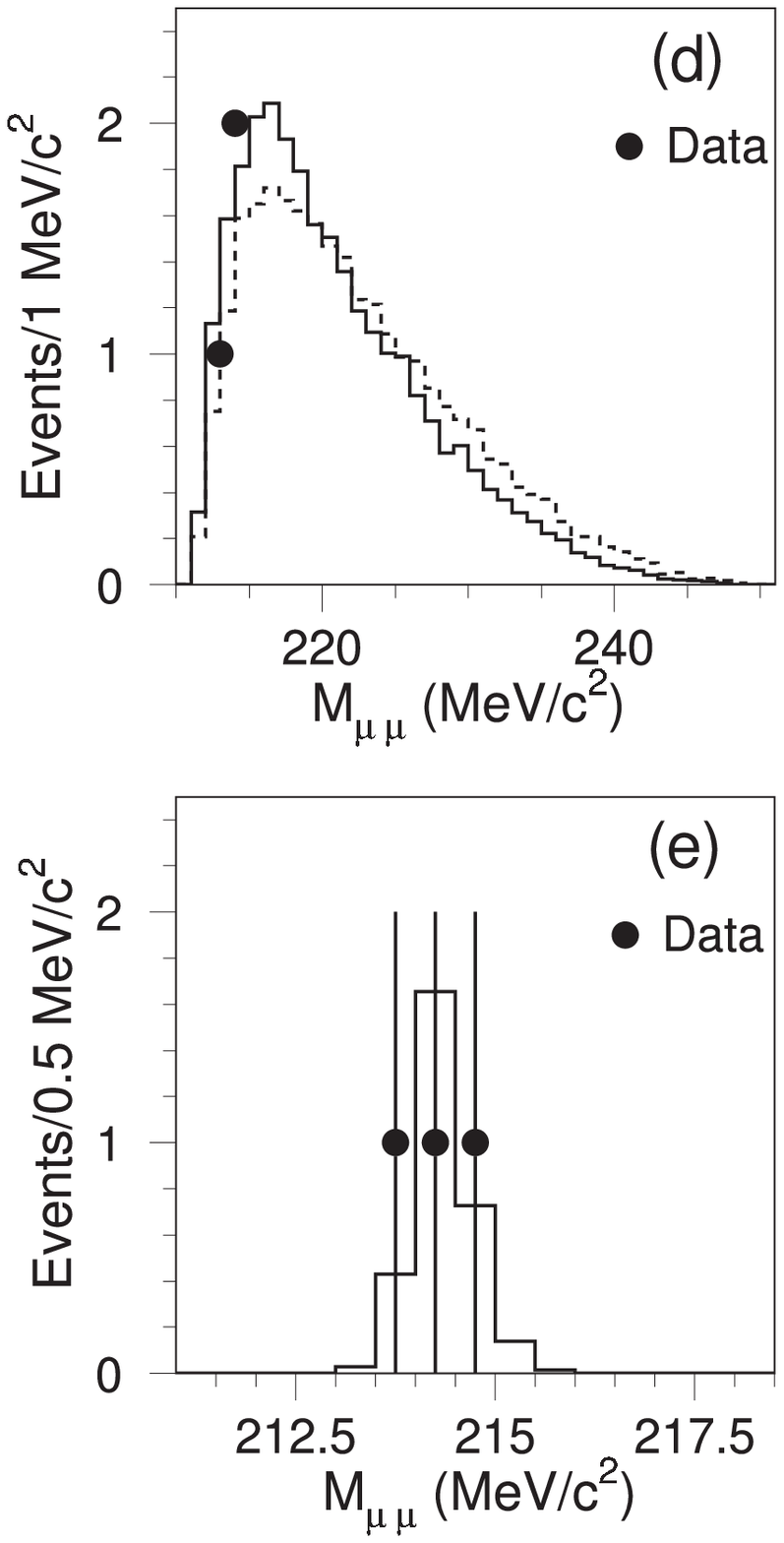}}
\end{minipage}
\caption{Mass spectrum for single-vertex $p\mu^+\mu^-$  candidates in HyperCP positive-beam data sample: a) wide mass range (semilog scale); b) narrow range around $\Sigma^+$ mass; c) wide mass range after application of additional cuts as described in Ref.~\protect\cite{Park-Sigpmumu}; dimuon mass spectrum of the three $\Sigma^+\to p\mu^+\mu^-$ candidate events compared with Monte Carlo spectrum assuming d) Standard Model virtual-photon form factor (solid) or isotropic decay (dashed), or e) decay via a narrow resonance $X^0$.}\label{fig:mumu}

\end{figure}

In addition to its high-rate charged-particle spectrometer, HyperCP had a muon detection system aimed at studying rare decays of hyperons and charged kaons~\cite{Park-Kpimumu,Park-Sigpmumu,Burnstein}. 
A recent HyperCP result is the observation of the rarest hyperon decay ever, $\Sigma^+\to p\mu^+\mu^-$~\cite{Park-Sigpmumu}. As shown in Fig.~ \ref{fig:mumu}, based on the 3 observed events, the decay is consistent with being two-body, i.e., $\Sigma^+\to p X^0,\,X^0\to\mu^+\mu^-$, with branching ratio ${\cal B}=(3.1^{+2.4}_{-1.9}\pm1.5)\times10^{-8}$ and $X^0$ mass $m_{X^0}=(214.3\pm0.5)\,$MeV. With the available statistics this interpretation is of course not definitive: the probability that the 3 signal events are consistent with the form-factor decay spectrum of Fig.~\ref{fig:mumu}d is estimated as 0.8\%; in this interpretation the measured branching ratio is ${\cal B}(\Sigma^+\to p\mu^+\mu^-)=(8.6^{+6.6}_{-5.4}\pm5.5)\times10^{-8}$.

This result is particularly intriguing given predictions by D. S. Gorbunov {\it et al.}~\cite{Gorbunov} of a pair of SUSY ``sgoldstino'' states (supersymmetric partners of Goldstone fermions). These can be scalar or pseudoscalar and they could be low in mass. It is thus conceivable that the lightest supersymmetric particle has now been glimpsed\,---\,and in a most unexpected place! This result demands further experimental study. But note that the HyperCP $\Sigma^+$ sensitivity was $\approx 2\times10^{10}$ decays. No planned experiment is capable of producing and detecting the ${\cal O} (10^{11})$ $\Sigma^+$ hyperons that would be required to confirm or refute HyperCP's putative $X^0$ signal.

\section{A future  experiment}

The Fermilab Antiproton Source is a suitable venue for a high-sensitivity experiment studying ({\it inter alia}) ${\overline p}p\to$\,hyperons.\footnote{Given the high rate of secondary beam in the HyperCP MWPCs ---
about 13\,MHz spread over an area of several cm$^2$ --- the HyperCP approach could not be pushed much further even were Tevatron fixed-target operation once again to be made available.} 
An appropriate goal would be an order-of-magnitude increase in sensitivity\,---\,two
orders of magnitude increase in sample size for the {\em CP} study and one order of magnitude for the rare-decay search\,---\,i.e., $\sim10^{11}$ ${\overline\Lambda}\Lambda$  and ${\overline \Sigma}\Sigma$ events. Cleanliness of the produced samples suggests operating just above threshold as in PS185~\cite{PS185}.  Given the $\approx60\,\mu$b cross sections~\cite{PS185}, $10^{33}\,{\rm cm}^{-2}{\rm s}^{-1}$ luminosity means $6\times10^{11}$ events produced per $10^7$-s year (which, with the inevitable acceptance and efficiency losses, probably
reaches the goal). A proto-collaboration is now forming to design and simulate the apparatus and produce a Letter of Intent.

\section{Acknowledgments}

I thank the organizers for the invitation to speak at this stimulating conference and C. Brown, D. Christian, K. Gollwitzer, G. Jackson, S. Nagaitsev, J. Ritman, and U. Wiedner for useful discussions.
This work was supported in part by a grant from the
U.S. Department of Energy.

\end{document}